%% file: ase23.tex
\documentclass[10pt,conference]{IEEEtran}
\IEEEoverridecommandlockouts
\usepackage[]{collab}
\usepackage{cite}
\usepackage{amsmath,amssymb,amsfonts}
\usepackage{algorithmic}
\usepackage{graphicx}
\usepackage{textcomp}
\usepackage{xcolor}
\usepackage{balance}
\usepackage{enumitem}
\usepackage{xspace}
\usepackage{subfigure}
\usepackage{booktabs}
\usepackage{multirow}
\usepackage[most]{tcolorbox}
\usepackage{makecell}
\usepackage[hidelinks]{hyperref}
\usepackage[ruled,linesnumbered]{algorithm2e}

\usepackage{titlesec}
\titlespacing\section{0pt}{6pt plus 4pt minus 2pt}{4pt plus 2pt minus 2pt}
\titlespacing\subsection{0pt}{6pt plus 3pt minus 2pt}{4pt plus 2pt minus 2pt}
\titlespacing\subsubsection{0pt}{6pt plus 3pt minus 2pt}{4pt plus 2pt minus 2pt}
\SetAlCapNameFnt{\footnotesize}
\SetAlCapFnt{\footnotesize}
\usepackage{amsmath}  
\makeatletter  
\g@addto@macro\normalsize{%  
\setlength\abovedisplayskip{0pt}  
\setlength\belowdisplayskip{0pt}  
\setlength\abovedisplayshortskip{0pt}  
\setlength\belowdisplayshortskip{0pt}  
}  
\SetKwInput{kwInit}{Init}

\def\BibTeX{{\rm B\kern-.05em{\sc i\kern-.025em b}\kern-.08em
    T\kern-.1667em\lower.7ex\hbox{E}\kern-.125emX}}

\definecolor{ballblue}{rgb}{0.13, 0.67, 0.8}
\collabAuthor{jy}{ballblue}{Jinyang Liu}
\collabAuthor{zh}{green}{Zhihan Jiang}
\collabAuthor{jz}{magenta}{Jiazhen Gu}
\collabAuthor{zb}{teal}{Zhuangbin Chen}

\newcommand\cluster{functional cluster\xspace}
\newcommand\clusters{functional clusters\xspace}
\newcommand\cloud{Huawei Cloud\xspace}
\newcommand\nm{Prism\xspace}
\newcommand\A{$\mathcal{A}$\xspace}
\newcommand\B{$\mathcal{B}$\xspace}

\newcommand{\ie}{{\em i.e.},\xspace}
\newcommand{\eg}{{\em e.g.},\xspace}

\begin{document}

\title{Prism: Revealing Hidden Functional Clusters from Massive Instances in Cloud Systems}

% \author{Anonymous author(s)}

\author{\large Jinyang Liu\IEEEauthorrefmark{2}$^*$, Zhihan Jiang\IEEEauthorrefmark{2}$^*$\thanks{\hspace{-2ex}$^*$Both authors make equal contribution to this paper.}, Jiazhen Gu\IEEEauthorrefmark{2}, Junjie Huang\IEEEauthorrefmark{2}, Zhuangbin Chen\IEEEauthorrefmark{3}$^{**}$\thanks{\hspace{-2ex}$^{**}$Zhuangbin Chen is the corresponding author.}, \\ Cong Feng\IEEEauthorrefmark{4}, Zengyin Yang\IEEEauthorrefmark{4}, Yongqiang Yang\IEEEauthorrefmark{4}, Michael R. Lyu\IEEEauthorrefmark{2}\\
\IEEEauthorblockA{
\IEEEauthorrefmark{2}The Chinese University of Hong Kong, Hong Kong SAR, China, \{jyliu, zhjiang22, jzgu, jjhuang23, lyu\}@cse.cuhk.edu.hk\\
\IEEEauthorrefmark{3}School of Software Engineering, Sun Yat-sen University, Zhuhai, China, chenzhb36@mail.sysu.edu.cn\\
\IEEEauthorrefmark{4}Computing and Networking Innovation Lab, Huawei Cloud Computing Technology Co., Ltd, China,\\ \{fengcong5, yangzengyin, yangyongqiang\}@huawei.com
}
}

\maketitle

\begin{abstract}
Ensuring the reliability of cloud systems is critical for both cloud vendors and customers.
Cloud systems often rely on virtualization techniques to create instances of hardware resources, such as virtual machines. 
However, virtualization hinders the observability of cloud systems, making it challenging to diagnose platform-level issues.
To improve system observability, we propose to infer \textit{functional clusters} of instances, \ie groups of instances having similar functionalities.
We first conduct a pilot study on a large-scale cloud system, \ie \cloud, demonstrating that instances having similar functionalities share similar \textit{communication} and \textit{resource usage} patterns. 
Motivated by these findings, we formulate the identification of functional clusters as a clustering problem and propose a non-intrusive solution called \textit{\nm}.
\nm adopts a coarse-to-fine clustering strategy. It first partitions instances into coarse-grained chunks based on communication patterns.
Within each chunk, \nm further groups instances with similar resource usage patterns to produce fine-grained functional clusters.
Such a design reduces noises in the data and allows \nm to process massive instances efficiently.
We evaluate \nm on two datasets collected from the real-world production environment of \cloud.
Our experiments show that \nm achieves a v-measure of $\sim$0.95, surpassing existing state-of-the-art solutions. Additionally, we illustrate the integration of \nm within monitoring systems for enhanced cloud reliability through two real-world use cases.

\end{abstract}

% \begin{IEEEkeywords}
% component, formatting, style, styling, insert
% \end{IEEEkeywords}

\input{content/01_introduction}

\input{content/02_background}

\input{content/03_method}

\input{content/04_evaluation}

\input{content/05_casestudy}

\input{content/06_discussion}
\input{content/07_relatedwork}

\input{content/08_conclusion}
\input{content/ack}

\balance
\bibliographystyle{IEEEtran}
\bibliography{ase23}

\end{document}

%% file: content/01_introduction.tex
\section{introduction}
% \jy{go through all terminologies}
% \jy{add JC magic code}
% Cloud computing has seen a significant increase in adoption by users worldwide in recent years.
% Large-scale cloud vendors, such as Amazon AWS~\cite{aws}, Microsoft Azure~\cite{azure}, and Google Cloud Platform (GCP)~\cite{gcp}, provide various types of services (\eg IaaS, PaaS and SaaS) in a 7$\times$24 manner to worldwide customers.
Cloud providers such as Amazon AWS, Microsoft Azure, and Google Cloud Platform (GCP) have provided a wide range of services and ensure availability 24/7 to their customers worldwide.
% large scale
Guaranteeing the reliability of a cloud system is crucial since even a brief downtime could result in significant financial losses for cloud vendors and their customers~\cite{coststudy, chen2019empirical}.

% problem, virtualization reduce observability -> challenging to ensure reliability
Cloud systems typically leverage virtualization techniques to abstract hardware resources, such as computation, storage, and networks, into instances (\eg virtual machines), serving as basic components of cloud services~\cite{jain2016overview, xing2012virtualization, malhotra2014virtualization}.
Such architecture provides flexibility and elasticity for tenants to subscribe various instances to run services with different functionalities \eg machine learning and database services. 
This, in turn, enables them to create complex and customizable applications.

However, just as each coin has two sides,
such practice makes it more challenging to ensure the reliability of cloud systems.
In particular, virtualization degrades the observability of the system, \ie the ability to understand the system internal execution state.
Virtualization introduces an additional layer of abstraction between the underlying hardware and the running applications, making it difficult to correlate the problems across different layers~\cite{wang2021fast}.
For example, an issue at the application layer may be caused by problems either within the instance itself or with the underlying hardware. 

% Existing methods and their limitations
% Since cloud vendors cannot touch tenants' private data, 
To enhance system observability, a common practice for cloud vendors is deploying a variety of monitors to collect runtime information of each instance~\cite{chen2020towards,chen2020incidental,ghosh2022fight,liu2023incident}, which record only data related to reliability issues without touching users' privacy.
The monitoring data are then utilized for downstream maintenance tasks. 
For example, \textit{communication traces}, which record network packet transmissions between instances (\eg the source and destination IP addresses and port numbers), are often used to identify abnormal network behaviors, such as network attacks and excessive traffics~\cite{iliofotou2007network, jin2009unveiling, nagaraja2010botgrep, xu2011network, arzani2020privateeye}.
On the other hand, \textit{performance metrics}, such as CPU utilization and memory usage, are commonly utilized for detecting anomalies and localizing faults~\cite{chen2022adaptive, zhao2021predicting, yang2021aid}.

The monitoring data have provided valuable insights to ensure the reliability of individual instances.
However, cloud vendors still view instances as distributed black boxes without knowing how an application is deployed across the infrastructure~\cite{pang2022cloudcluster}.
Consequently, it can be challenging to assess the impact of issues at the platform level (such as instance or hardware problems) on applications that are deployed on top of them. 
For example, packet losses in individual instances are commonplace in cloud systems and are generally ignored, as they seldom impact customer applications.
However, when multiple instances, all supporting the same application, concurrently experience packet losses, it likely indicates a more significant issue that users may encounter, such as interruptions due to network disconnections.
The limited awareness of relationships between instances complicates the detection of such problems, thereby impeding timely mitigation efforts.

To bridge this gap and improve the system observability, we propose to infer \textit{functional clusters} of instances, where each cluster contains the instances having similar functionalities.
With this additional knowledge, cloud vendors can enhance the reliability of the cloud by improving various downstream management and maintenance tasks (to be detailed in \S \ref{sec: case_study}). 
However, there are two major challenges that we need to overcome to achieve this goal.
The first challenge is that only limited information is available. As mentioned before, cloud vendors cannot access tenants' private data, including logs and source codes. A non-intrusive solution that relies solely on external data (\eg traces and metrics) is required.
The second challenge is the large scale of instances in cloud systems.
A typical cloud system can consist of millions of instances in total~\cite{pang2022cloudcluster}, resulting in an enormous amount of data for analysis.
Valuable insights are concealed within the vast and noisy data of cloud systems, making it difficult to reveal the hidden function clusters.

% empirical study: two similarities
To tackle the first challenge and explore a feasible non-intrusive solution, we first conduct a pilot study on the services deployed in \cloud.
For privacy reasons, we only use internal services of \cloud without touching tenants' instances.
Specifically, we utilize a total of 3,062 internal instances covering services with 397 types of functionalities and study whether different functionalities can be identified simply based on external monitoring data.
Our study uncovers that instances having similar functionalities share similar \textit{communication} and \textit{resource usage} patterns.
Communication patterns mean that instances with similar purposes may frequently communicate with the same set of destinations, reflected in their communication traces.
We find that 75\% of instances within the same \clusters have a high overlap ($\geq$ 0.7) in their communicated destinations. Conversely, for 92\% of instances with different functionalities, the overlap is only less than 0.2.
Additionally, despite the large scale of instances, 99.1\% of instances communicated with a limited number of destinations (fewer than 50), indicating a strong locality in communication patterns.
% similar kpi 
Resource usage patterns, on the other hand, denote that instances with similar functionalities would demonstrate comparable resource consumption, which is reflected in their metrics.
For example, a machine learning service is expected to exhibit greater CPU usage, while instances running an in-memory database like Redis would primarily require more memory.
We find that most ($\sim$75\%) of instance pairs with the same functionalities have high metric-based similarities ($\geq$ 0.8), while the similarities decrease for those instances having different functionalities.
% similar communication

Motivated by the two kinds of inherent patterns of the instances, we formulate the identification of \clusters as a clustering problem. 
Intuitively, we aim to cluster the instances by harmoniously integrating the communication patterns and resource usage patterns. 
To achieve this goal and alleviate noises within the tremendous data, we propose \textit{\nm}, which adopts a coarse-to-fine clustering strategy.
\nm consists of two components, \ie \textit{trace-based partitioning} and \textit{metric-based clustering}.
In the trace-based partitioning step, we leverage the communication patterns to coarsely divide the entire large set of instances into smaller chunks.
This step helps limit the comparison space within each chunk, thus reducing the complexity of the subsequent clustering process and eliminating noises introduced by instances from other clusters.
In the metric-based clustering step, we perform fine-grained clustering by comparing the resource usage patterns of instances in a pairwise manner. 
This step allows us to carefully group instances within the same \cluster.

To evaluate \nm, we conduct extensive experiments on two datasets collected from the production environment of \cloud, a top-tier cloud provider serving global customers. 
To evaluate the generality of \nm, these datasets were procured from two regions of \cloud, each covering a diverse set of functionalities.
The experimental results show that \nm achieves a v-measure of $\sim$0.95, surpassing existing state-of-the-art solutions, and is robust to parameter  changes. 
Moreover, \nm is both scalable and efficient, with a linear time complexity, enabling it to handle a substantial number of instances.
Furthermore, we have deployed \nm in \cloud, and we share two real-world use cases to demonstrate the usefulness of functional clusters in maintaining \cloud.
% The first scenario highlights the ability of \clusters to identify vulnerable deployments of applications that may be prone to disruption from hardware failures.
In the first case, \clusters showcase the ability to detect vulnerable application deployments that may be at risk of disruption due to hardware failures.
The second case shows how \clusters can aggregate minor packet loss errors across instances, thus enabling identification of latent issues that are not observable at either the instance or region level.
We summarize our contributions as follows:
\begin{itemize}[leftmargin=*, topsep=0pt]
    \item We conduct a pilot study to understand the characteristics of \clusters across over 3,000 instances based on a real-world cloud system, \cloud (\S\ref{sec: background_and_motivation}). Our findings reveal two clues for identifying \clusters (\ie communication patterns and resource usage patterns).
    
    \item We design a non-intrusive solution called \nm to identify \clusters in large-scale cloud systems, which is able to effectively capture and integrate the inherent communication and resource usage patterns among instances (\S\ref{sec: method}).
    
    \item Extensive experiments are conducted on two real-world industrial datasets (\S\ref{sec: evaluation}).
    Our results demonstrate that \nm is effective, efficient and practically useful in identifying \clusters in industrial cloud systems.
    Our dataset and code are made public to benefit the community on \url{https://github.com/OpsPAI/Prism}.
\end{itemize}
% 

% \begin{figure*}[t]
%   \centering
% \mbox{
% %   Impact of alerts with different levels of severity
%      \subfigure[\label{fig: resource_pattern}]
%      {\includegraphics[width=0.6\columnwidth]{figures/resource_pattern.pdf}} \quad
    
%     \subfigure[\label{fig: communication_pattern}]
%     {\includegraphics[width=0.6\columnwidth]{figures/communication_pattern.pdf}} \quad

%      \subfigure[\label{fig: destination}]
%     {\includegraphics[width=0.603\columnwidth]{figures/destination.pdf}} \quad
% }
    
%   \caption{ \cloud}
% \end{figure*}
% exising solutions for observerbility

% \begin{itemize}
%     \item cloud systems serve 7x24h service
%     \item cloud provides virtualized instances to customers for (1) maximum resource utilization, (2) customers' easy use
%     \item  
% \end{itemize}

% \begin{itemize}
%      \item such virtualization poses challenges for cloud observerbility for ensuring reliability: hard to localize -> customers' application running on VMs -> 
%     \item 
% \end{itemize}

%% file: content/02_background.tex
\begin{figure}[t]
    \centering
    \includegraphics[width=0.95\linewidth]{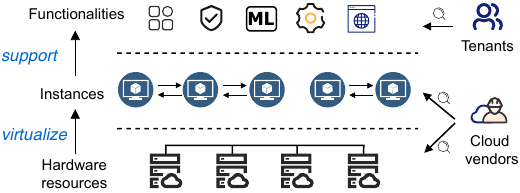}
    \vspace{-5pt}
    \caption{The hierarchical structure of cloud systems}
    \label{fig: cloud_structure}
\end{figure}
\begin{figure*}[t]
    \centering
    \includegraphics[width=1.9\columnwidth]{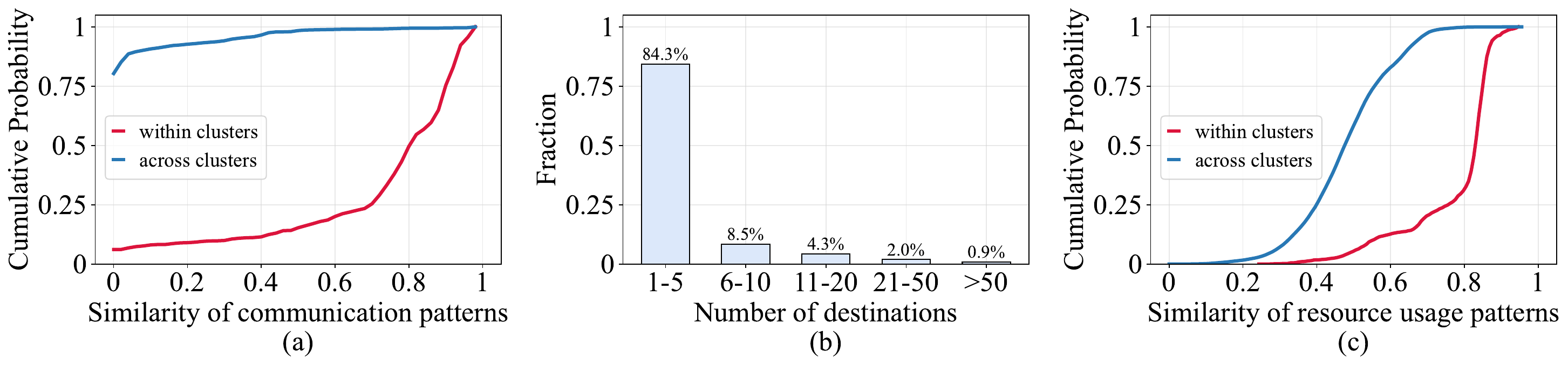}
    \vspace{-10pt}
    \caption{Results of the study on communication and resource usage patterns.}
    \label{fig: pattern_study}
    \vspace{-18pt}
\end{figure*}

\section{Background and Pilot Study}\label{sec: background_and_motivation}
% \jz{this section should be reorganized, see anotations}
% background
% 1. current system structure: application, service, instance
% 2. external monitoring(performance metrics, traffic)
% 
% pilot study
% 1. functional clusters (instances running similar servicers) definition  
% 2. external metric similarity of functional clusters

In this section, we first discuss the background of cloud systems and clarify the terminologies used. 
Then, we present a pilot study to understand the characteristics of instances that can facilitate the identification of \clusters.

\subsection{Background}

\subsubsection{Cloud System Structure}
Modern cloud systems are complex and highly distributed, consisting of multiple layers of hardware and software components that work together to provide on-demand computing resources to tenants. 
Fig.~\ref{fig: cloud_structure} shows the hierarchical structure of a typical cloud system.
At the lowest layer, hardware resources such as physical machines, storage, and networking equipment form the underlying infrastructure.
These resources are virtualized into environments known as \textit{instances} (\eg virtual machines), which can be dynamically created, scaled, and terminated as needed, forming a layer of virtualization.
On top of these instances, tenants can deploy services with a broad range of \textit{functionalities} that run in different programming languages and frameworks. 
These functionalities can either serve as standalone \textit{applications} or be combined to form complex \textit{applications}. 
For example, an online shopping application may consist of services offering interdependent functionalities like load balancing, user authentication and databases. 
To ensure scalability and fault tolerance, multiple copies (or replicas) of the same service are typically created and distributed across the cloud environment to support a single functionality. 
This approach enables handling user traffic spikes while guaranteeing high availability of the system in case of instance failures. 
It is crucial to timely detect potential issues in the services with various functionalities that constitute the application in order to ensure its overall availability.
% \zb{It is crucial to timely detect potential issues in all services of the application, thereby ensuring the availability of its functionalities utilized by customers.}
This paper focuses on discovering functional clusters containing instances with similar functionalities, which are smaller and more manageable units than complex applications.
With this information, operators are allowed to build more actionable monitoring metrics for safeguarding each functionality.

\subsubsection{Cloud System Monitoring}
% \jz{above is duplicated with the introduction, can simplify the introduction part.}
Monitoring is a common practice for top-tier cloud vendors, such as  AWS CloudWatch~\cite{aws_cloudwatch}, Azure Monitor~\cite{azure_monitor} and Google Cloud Monitoring~\cite{gcp_monitoring}.
Monitoring tools are used to collect various types of data about the system's performance and behavior. 
Two key types of data are commonly collected for each instance: communication traces and performance metrics.
\textit{Communication traces} data, are records of network transmissions between instances in a cloud environment. 
These traces are typically generated by network monitoring tools (\eg flow logs~\cite{gcp_flowlog, aws_flowlog, azure_flowlog}) and capture metadata about the network traffic, such as the source and destination IP addresses, port numbers, and protocol types.
By collecting and analyzing these types of data, cloud system operators can take proactive measures to ensure system security and reliability~\cite{iliofotou2007network, jin2009unveiling, nagaraja2010botgrep, xu2011network, arzani2020privateeye}.
\textit{Performance metrics}, on the other hand, includes information such as memory usage and network throughput organized in the form of time series, which are used to detect and diagnose system performance issues ~\cite{chen2022adaptive,zhao2021predicting}.

As shown in Fig.~\ref{fig: cloud_structure}, tenants mostly focus on the functionalities of services they deploy, rather than delving into infrastructure-level details.
On the other hand, due to privacy concerns, cloud vendors only possess runtime information about instances and hardware resources, lacking knowledge about how customers deploy services with various functionalities across these instances.
While cloud providers do possess some metadata about these instances, such as which customers subscribe to particular instances, this information is often too coarse-grained. For example, thousands of instances belonging to the same enterprise customer could share the same tenant ID~\cite{pang2022cloudcluster}, and these instances may host a diverse range of functionalities.
This paper aims to empower cloud providers with insights into more fine-grained structure of the functionalities by learning from instance data visible to them.
This would facilitate building an enhanced monitoring system to improve the reliability of cloud systems (will show in \S\ref{sec: case_study}).

\begin{figure*}[t]
    \centering
    \includegraphics[width=1.9\columnwidth]{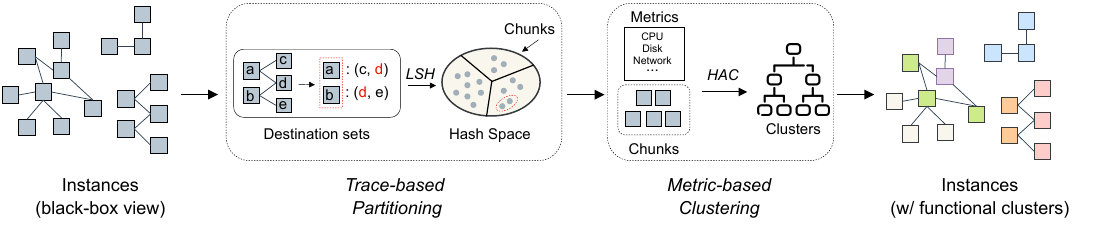}
    \vspace{-8pt}
    \caption{The overall workflow of \nm}
    \label{fig: method_framework}
    \vspace{-20pt}
\end{figure*}

\subsection{A Pilot Study}
% \jy{or problem statements}
% \zb{I think you need more words to clearly introduce what the \cluster of instances is and what we can do with them.}
% % what are instances
% % why there are \clusters
% % what is the importance of the awareness of the \clusters
% We define \clusters as clusters of instances in which each cluster comprises instances performing similar tasks, such as a cluster of instances running a database or Nginx service.
% % Virtualization of resources enables cloud providers to scale resources up or down easily and quickly based on customer demand. 
% To ensure high availability and scalability~\cite{gunawi2016does}\cite{lin2018predicting}, an application can involve many instances, which can work as replicas for fault tolerance or load balancers to help balance the workload of other instances.
% With such awareness of \clusters across the massive instances in a cloud system, cloud vendors can make better decisions for optimizing the instance deployment, improving the diagnosis procedure and configuring more fine-grained monitoring (as will show in Section~\ref{sec: rq4_usefulness})\jy{final check here}.
In the following, we conduct a pilot study across over \textit{three thousand} internal instances in \cloud, aiming to find clues to uncover the valuable \clusters.
We conduct manual inspections in collaboration with the corresponding teams within \cloud to understand their functionalities. We obtain services covering 397 types of functionalities in total, and more details about this dataset are in \S\ref{sec: exp_setup}.

\subsubsection{Communication Pattern}\label{sec: communication_pattern_study}
The communication pattern serves as an indicator that instances within the same \cluster tend to exhibit comparable network behaviors, as evidenced by the communication traces they generate.
% Communication trace is another type of monitoring data available to cloud vendors.
% They comprise records of network transmissions from an instance (\ie source) to other internal instances or external hosts (\ie destination).
% Typically, cloud vendors such as Google Cloud Platform~\cite{pang2022cloudcluster} and Microsoft Azure~\cite{roy2018cloud} collect these records to assess network health, and these communication traces only carry network packets (\ie binary bits) without any human-readable information.
As inspired by~\cite{pang2022cloudcluster}, instances within the same \clusters might communicate with similar destinations.
To investigate this, we combine every two instances and compute the overlap of their destinations through Jaccard similarity~\cite{jaccard}. 
% The calculation methodology is presented in detail in \S\ref{sec: method}.
Then, we compare the similarities within the same clusters and across different clusters.

Fig.~\ref{fig: pattern_study}-(a) presents the comparison results of communication pattern similarities within or across clusters, where we can observe a significant difference between them.
When examining instances within the same cluster, we find that 50\% of the instance pairs demonstrate more than 0.8 similarity and over 75\% of them exhibit more than 0.6 similarity. In contrast, when comparing instance pairs from different clusters, over 75\% of the pairs exhibit a similarity score of 0, indicating no overlap between their destinations. Additionally, 96\% of the pairs have a similarity score of $<$0.4, indicating that instances from different clusters rarely communicate with the same destinations.
However, for some cross-cluster instances, there are still a little overlap in their destinations. 
These destinations are usually common services such as network gateway and authentication that are shared by multiple applications.

To further understand the communication patterns, we study how many different destinations one instance can frequently communicate with.
Fig.~\ref{fig: pattern_study}-(b) shows the results. We can find that even though there are thousands of instances in total, the majority of the instances only communicate with a small number of destinations. For example, 84.3\% of instances communicate with 1 to 5 instances, and 99.1\% of instances communicate with less than 50 instances.
This suggests a strong \textit{locality} of instances, \ie most instances tend to communicate with a small set of other instances frequently.

% \subsubsection{Findings}
% {
% \begin{tcolorbox}[breakable,width=\linewidth-2pt,boxrule=0pt,top=1pt, bottom=0pt, left=1pt,right=1pt, colback=gray!20,colframe=gray!20]
% \textbf{Summary of findings}

% \end{tcolorbox}
% }

\subsubsection{Resource Usage Pattern}\label{sec: resource_pattern_study}
Intuitively, instances within the same \cluster should observe similar patterns in their resource consumption (\ie resource usage patterns). 
% To ensure the reliability of each instance, cloud vendors generally configure monitors to continuously collect metrics (\eg CPU/memory utilization and disk read/write rate) at fixed time intervals, such as every five minutes.
To investigate whether resource usage patterns can be utilized to uncover \clusters, we analyze the similarities in the metric data among instances, either within the same \cluster or across different clusters. 
Thus, we compare the multivariate metric similarity on two instances using the multivariate dynamic time warping (DTW) distances~\cite{muller2007dynamic}, a distance metric to compare a pair of time series that may vary in timing (more details in \S\ref{sec: method}).

Fig.~\ref{fig: pattern_study}-(c) shows the distribution of resource usage pattern similarities among instances, either within or across \clusters. 
We can observe that the similarities of instance pairs within clusters are generally large, with over 75\% of such pairs exhibiting 0.7 similarity or higher.
In contrast, instance pairs across clusters display smaller similarities, with 92\% of pairs across different clusters possessing less than 0.2 similarity. 
However, it is worth noting that there is a small portion ($\leq$10\%) of cross-cluster instance pairs that have high metric-based similarities, with a value of $\geq$0.8.
This is reasonable since instances having different functionalities could behave similarly, \eg have a high CPU utilization.
Nevertheless, it still suggests that leveraging the similarities between instance metrics is promising in distinguishing their \clusters.

% {
% \begin{tcolorbox}[breakable,width=\linewidth-2pt,boxrule=0pt,top=1pt, bottom=0pt, left=1pt,right=1pt, colback=gray!20,colframe=gray!20]
% \textbf{Finding 1} 
% Instances performing similar tasks tend to possess similar consumption of resources, which is reflected by high similarities between their metrics, while instances across different \clusters are less similar in their metrics.
% \end{tcolorbox}
% }

\noindent\textbf{Summary.} We summarize our findings as follows.

\begin{itemize}[leftmargin=*, topsep=0pt]
    \item Instances that belong to the same \cluster exhibit comparable communication patterns, as evidenced by the considerable overlap in their communication destinations.
    Furthermore, the analysis reveals that the majority of instances interacted with a limited number of other instances, indicating a strong locality of instances.
    \item Instances within clusters generally exhibit high similarities in their resource usage patterns, while instance pairs across clusters show smaller similarities.
    \item While communication and resource usage patterns provide valuable insights, they are not entirely reliable indicators for distinguishing between different \clusters, as some \textit{noises} in the form of cross-cluster instances with high similarities in both patterns are observed. % \jz{this challenge is addressed by the coarse-to-grain strategy?}\jy{yes, will mention}.
\end{itemize}
 
% These findings motivate us to develop an effective solution for discovering \clusters that can harness the benefits of both communication patterns and resource usage patterns while mitigating the inherent noises in both data.

%% file: content/03_method.tex
\section{Methodology}\label{sec: method}
% \subsection{Problem Statement}

% In doing this, we derive high-level application information 

\subsection{Overview}
The goal of this paper is to design a non-intrusive solution to discover functional clusters among massive instances in a large-scale cloud system. 
The input is an entire set of instances and their associated monitoring data, \ie communication traces and performance metrics.
The output of our approach is multiple clusters, where each cluster represents a functional cluster consisting of instances that have similar functionalities.

To achieve this goal, we propose \nm, an automated approach that can effectively discover \clusters based on both the communication patterns and resource usage of instances. 
Fig.~\ref{fig: method_framework} illustrates the overall workflow of \nm, which comprises two main components: \textit{trace-based partitioning} and \textit{metric-based clustering}.
Given a set of instances, \nm adopts a two-stage clustering process, which progressively divides the entire set of instances to coarse-grained \textit{chunks}, then fine-grained \textit{functional clusters}.
Specifically, the \textit{trace-based partitioning} step is inspired by the strong locality of communication patterns, as shown in \S\ref{sec: communication_pattern_study}. 
Based on communication patterns, \nm first separates all instances into different chunks. Instances in the same chunk share similar communication destinations.
By dividing the complete instances set into multiple small chunks, we can reduce the noises introduced from other instances during the subsequent fine-grained clustering step.
For each chunk, \textit{metric-based clustering} is then applied to generate fine-grained clusters by measuring the similarities of monitoring metrics of instances. Finally, instances belonging to the same resultant cluster are considered to have similar functionalities.
Such a coarse-to-fine design avoids pairwise comparisons between a large number of instances and reduces noises between instances, making \nm salable and practical for large-scale cloud systems.

% incremental insertion
% It is worth noting that \nm only uses external monitoring data without touching tenants' data, and thus raising no privacy issue.
% With \nm, we can only infer which instances have similar functionalities while cannot identify the specific functionalities in use, thereby ensuring tenants' confidentiality.
It is important to note that \nm relies solely on external monitoring data and does not access any of the tenants' private data, which ensures that there are no privacy concerns.
While we can infer which instances have similar functionalities, we cannot identify the specific type of the functionalities in use. This approach maintains our tenants' confidentiality.

\subsection{Trace-based partitioning}\label{sec: trace_based_partition}
% rationale
As studied in \S\ref{sec: communication_pattern_study}, instances sharing the same \clusters are more likely to communicate with a similar set of destination hosts.
Thus, the trace-based partitioning of \nm measures the communication pattern similarity and divides instances into coarse-grained \textit{chunks}.

\noindent\textbf{Data Preprocessing.}
Let $x_i$ represent an instance in the cloud system. 
Communication traces can be represented as tuples of the form ($x_{src}$, $x_{dst}$), where $x_{src}$ and $x_{dst}$ represent the instances that communicate with each other. 
By analyzing the communication traces, we can obtain the \textit{destination set} of each instance, denoted by $S_i$ = ($x_1, x_2, x_3, ...$), which contains all the instances that have communicated with $x_i$.
However, as demonstrated in \S\ref{sec: communication_pattern_study}, instances with dissimilar functionalities may share common destinations, such as network gateways, which can introduce noise when comparing the communication patterns between instances. 
To mitigate this issue, we remove instances that interact with more than 100 different instances, which is rare as shown in Fig.~\ref{fig: pattern_study}-(b).

\noindent\textbf{Jaccard Similarity-based Partitioning.}
Next, we divide all instances into chunks by measuring how much their destination sets overlap.
To achieve this, a straightforward solution is to calculate the Jaccard similarity~\cite{tanimoto1958elementary} of destination sets of every pair of instances, which is denoted as $J(x_i,x_j) = \frac{|S_i \cap S_j|}{|S_i \cup S_j|}$, \ie the ratio of the size of their intersection to the size of their union.
However, it requires conducting pairwise comparisons between millions of instances in a large-scale cloud system.
This process can be extremely time-consuming and may render the approach unfeasible in practice.
 
To address this issue, we propose to leverage locality-sensitive hashing (LSH)~\cite{leskovec2020mining} to enable efficient partitioning.
LSH is a technique developed for identifying similar items in large datasets.
Its idea involves hashing the items into signatures such that similar items are more likely to be assigned to the same bucket. 
Given a query, LSH can efficiently return similar items with a sub-linear time cost without pairwise comparison with the entire instance set.
In our context, we combine LSH with the MinHash function, which allows items with high Jaccard similarities put into the same buckets~\cite{broder1997resemblance}.

Algorithm~\ref{algo: trace_based_partition} describes the trace-based partitioning process.
First, we extract the destination sets $S$ of each instance from historical communication traces (lines 1-5).
Second, for each instance $x_i$, we apply MinHash function to its destination set $S_i$ to obtain the hash signature. The hash signature is then inserted into the LSH model (lines 7-10), which assigns the item to a bucket.
Third, for each instance $x_i$, we search its nearest neighbors $\mathcal{N}_i$ within the buckets produced by the LSH model (lines 12-14). Here, a manual-defined threshold $\theta_{LSH}\in [0,1]$ is included, where a smaller $\theta_{LSH}$ value allows more dissimilar neighbors to be included.
After that, we group the instance $x_i$ with its neighbors $\mathcal{N}_i$ based on the Disjoint-set data structure $U$ (lines 15-19).
This data structure $U$ provides two efficient operations, \ie $U.findSet$ that find the set that contains a specific item and $U.unionSet$ that merge two disjoint sets.
If we find the sets containing $x_i$ and containing $x_j$ are disjoint (line 16), we merge these two sets (line 17) since $x_i$ and $x_j$ are similar.
In this way, we progressively divide the entire set of instances into multiple disjoint sets (\ie chunks) managed by $U$. 
Finally, we can obtain all the instance chunks $\mathcal{C}$ by enumerating the records in $U$ (line 21).

\input{algs/trace_partition}

The trace-based partitioning algorithm is highly efficient for two reasons.
First, we bypass the expensive pairwise similarity computation for all the instances by using LSH with MinHash. 
Secondly, we leverage the disjoint-set data structure to merge similar instances into chunks efficiently. The \textit{findSet} and \textit{unionSet} operations of the disjoint-set data structure can be completed within nearly constant time complexity, which further ensures the efficiency of the merging process.
Moreover, the number of neighbors $\mathcal{N}_i$ (line 14) is generally fewer than 50, which is much smaller than the total instance number $\mathcal{X}$ (line 12) due to the locality of communication patterns (Fig.~\ref{fig: pattern_study}-(b)), which improves \nm's scalability, making it feasible for large-scale cloud systems like \cloud.

% LSH enables us to retrieve those items that have a Jaccard similarity greater than a threshold $\theta_{LSH}$ in linear time complexity. 
% In doing this, we can efficiently obtain multiple instance chunks, where each chunk contains instances with destination sets that preserve an overlap to some degree, as determined by $\theta_{LSH}$.
% \jy{explain LSH sacrifices a little precision for high efficiency.}

% \begin{table}[t]
% \centering
% \caption{Monitoring Metrics for Clustering}
% \label{tab: term_definition}
% \begin{tabular}{cc}
% \hline
% \textbf{Metric} & \textbf{Decription} \\ \hline
% CPU Utilization\centering & \makecell*[l]{An alert is triggered when abnormal behavior of\\ a component is detected.}\\ \hline
% Ticket\centering & \makecell*[l]{A request raised by a customer to ask the cloud vendor\\ for help.} \\ \hline
% Incident\centering & \makecell*[l]{Unexpected interruptions affecting services' availability\\ or performance, which usually trigger a series of alerts.} \\ \hline
% \makecell*[c]{Alert-Alert\\Relation} & \makecell*[l]{Two alerts are correlated  if they are caused by the\\ same incident. (Section~\ref{sec: alert_alert_relation})}\\\hline
% \makecell*[c]{Ticket-Alert\\Relation} & \makecell*[l]{A ticket is correlated with an alert if the former is\\ caused by the latter. (Section~\ref{sec: alert_ticket_relation})}\\
% \hline
% \end{tabular}
% \end{table}

\subsection{Metric-based Clustering}\label{sec: metric_based_clustering}
Trace-based partitioning tends to group as many instances as possible together, which can inevitably include instances with different functionalities to the same chunk. 
The reason is that instances from different clusters can still communicate to the same destinations (as studied in \S\ref{sec: communication_pattern_study}), and this leads to overlap of the destination sets of these instances, which may be wrongly grouped together.

To address this problem, we further group these instances by utilizing more fine-grained monitoring metrics that record detailed runtime information of instances (\ie resource usage patterns as studied in \S\ref{sec: resource_pattern_study}). 
Each instance is monitored via multiple dimensions to ensure its reliability, producing multivariate metrics, including CPU utilization rate, network incoming/outgoing bytes rate, disk read/write request rate, and disk read/write bytes rate.
In the following, we aim to calculate a metric-based distance for each pair of instances.
Then, we can cluster those instances that are close to each other.

\noindent\textbf{Data Preprocessing.}
We apply the following preprocessing techniques to the raw metric data collected to remove noises and normalize the data within a comparable scale.
First, we regard apparent extreme values as anomalous noises within the metric data because these values can bias the subsequent distance computation step.
For each metric, we replace the data points that are out of the three-sigma range with the average value of the nearest ten points. 
Next, since the amplitude scales of different metrics are different, \eg network-related metrics are highly variable and may range from tens of bytes to millions of bytes. This can make the produced distances incomparable between instances with different network traffic volumes. 
To address this issue, we apply natural logarithm to these metrics following~\cite{pang2022cloudcluster} to make it more robust to its variance.
The logarithm only solves the issue of highly variable amplitudes but does not ensure that the data points fall within the same range.
Therefore, finally, we apply min-max normalization to scale each of the metrics to the range of 0-1, allowing comparison across different metrics. Formally, using y to denote a metric time series, the normalized values can be calculated as $y'=\frac{y-min(y)}{max(y)-min(y)}$.

\noindent\textbf{Metric-based Distance Calculation.}
For an instance $x$, its preprocessed monitoring metrics form a group of multivariate time series represented as a matrix $\textbf{M}_i \in \mathbb{R}^{n\times k}$, where $n$ is the number of timestamps and $k$ is the number of metrics used.
We measure the metric-based similarity of two instances using a distance that simultaneously considers all the multivariate metrics of them.
To achieve this, we first compare each metric, then aggregate the distances to produce an overall distance.

Specifically, we adopt dynamic time warping (DTW) distances~\cite{muller2007dynamic} for distance measurement.
The reason we use DTW is to overcome the problem that the monitoring metrics of different instances can have time shifts, namely, these time series may not be aligned in terms of the collection timestamps, making traditional distance measures such as Euclidean distance ineffective.
In contrast, DTW allows for flexible matching of similar patterns in the time series, even when they occur at different timestamps.
Based on the DTW calculation, the overall distance $d(x_i, x_j)$ between two instances $x_i$ and $x_j$ can be formulated as follows:
\begin{align}
    d(x_i, x_j) &= \sum_{u=1}^k \omega(i,j)_u \times DTW\big( \textbf{M}_i(:,u), \textbf{M}_j(:,u) \big), \\
    \omega(i,j)_u &= \frac{\omega(i,j)_u'}{\sum_{v=1}^k \omega(i,j)_v'}, \label{equ: weight_normalize} \\
    \omega(i,j)_u' &= \frac{1}{2}\big( \sigma(\textbf{M}_i(:,u)) + \sigma(\textbf{M}_j(:,u))\big), \label{equ: average_std}
\end{align}

\noindent where $u$ denotes the metric in concern, $\textbf{M}_{i/j}(:.u)$ is the $u_{th}$ column of the corresponding metric matrix. 
In particular, we use $\omega(i,j)_u$ as a weight associated with the $u_{th}$ metric to measure the importance of each metric.
Each weight is calculated as the average of the standard deviation (\ie $\sigma(\cdot)$) of the two metrics of corresponding instances as shown in Equation~\ref{equ: average_std}, which is normalized to the range of 0 to 1 across different metrics using Equation~\ref{equ: weight_normalize}.
In doing this, we reduce the weight of the metrics that barely fluctuate (\eg two instances keep the CPU utilization rate around 80\%), since these metrics are less informative in representing the characteristics of instances.
In contrast, if two metrics are simultaneously changing following the same trend, they are more likely to indicate instances performing the same functionalities.

\noindent\textbf{Clustering Algorithm.}
We then apply a clustering algorithm in each \textit{chunk} based on the metric-based distances to produce more fine-grained clusters (\ie \clusters).
Specifically, we choose the hierarchical agglomerative clustering (HAC)~\cite{nielsen2016hierarchical} algorithm because it allows us to adjust the number of produced clusters via setting a distance threshold, \ie $\theta_{HAC}$.
The clustering algorithm starts by considering each instance as a single cluster and then iteratively merges the closest pairs of clusters until a user-defined threshold $\theta_{HAC}$ is reached. 
In this process, we use complete linkage~\cite{defays1977efficient} to find the closest pair of clusters, \ie the distance between two clusters is defined as the maximum DTW distance between any pair of instances in the two clusters.

While HAC requires the computation of distances between instances in a pairwise manner, it is still efficient since HAC is applied separately in each chunk.
Recall that chunks are produced by the trace-based partitioning step, and each chunk only contains tens of instances because of the locality of communication patterns (as shown in \S\ref{sec: communication_pattern_study}). 
Therefore, the computation within each small chunk can significantly reduce the computation cost, making our framework scalable to a large number of instances in cloud systems.
% \subsubsection{Incremental Insertion}

%% file: algs/trace_partition.tex
\begin{algorithm}[t]
\small
\caption{Trace-based Partitioning}
\label{algo: trace_based_partition}
% \normalsize
\SetAlgoLined
\KwIn{List of instances: $\mathcal{X}=\{x_1, x_2, ..., x_l\}$; Communication trace records: $\mathcal{R}=\{r_1, r_2, ..., r_t\}$; Similarity threshold: $\theta_{LSH}$}
\KwOut{Multiple instance chunks: $\mathcal{C}=\{C_1, C_2, ...\}$}

% pair-wise combination as a big weighted graph
\kwInit{$S$ $\leftarrow$ Empty list of feature sets; $M_{LSH}$ $\leftarrow$ empty LSH model; $U$ $\leftarrow$ Disjoint-set data structure}

// (1) Construct feature sets

\For{$i\leftarrow 1$ \KwTo $t$}{
    $x_{src}, x_{dst} \leftarrow r_{i}$
    
    $S[x_{src}]$.insert($x_{dst}$) 

}

// (2) Build the LSH model

\For{each instance $x_i \in \mathcal{X}$ }{
    $S_i\leftarrow S[x_i]$
    
    $M_{LSH}$.insert(MinHash($S_i$))
}

// (3) Search neighbors and build chunks

\For{each instance $x_i \in \mathcal{X}$ }{
    $S_i\leftarrow S[x_i]$ % already done by line 8?

    $\mathcal{N}_i$ = $M_{LSH}$.search($S_i$, $\theta_{LSH}$) // find neighbors

    \For{each instance $x_j \in \mathcal{N}_i$ }{
        
        \If{$U$.findSet($x_i$) != $U$.findSet($x_j$)} {
            $U$.unionSet($x_i$, $x_j$) // merge $x_i$ and neighbors
        }
        
    }
    
}

$\mathcal{C} \leftarrow$ U.getAllSets()
\vspace{-3pt}
\end{algorithm}

%% file: content/04_evaluation.tex
\section{Evaluation}\label{sec: evaluation}
We evaluate \nm by answering the following research questions (RQs):
\begin{itemize}[leftmargin=*, topsep=0pt]
    \item \textbf{RQ1:} How effective is \nm in clustering instances having similar functionalities?
    \item \textbf{RQ2:} How does each component contribute to the overall performance of \nm?
    \item \textbf{RQ3:} What is the parameter sensitivity of \nm?
    \item \textbf{RQ4:} What is the efficiency of \nm?
\end{itemize}

\begin{table}[]
\footnotesize
\centering
\caption{Dataset Statistics}\label{tab: dataset_statistics}
\vspace{-5pt}
\begin{tabular}{ccccc}
\toprule
 Datasets & \# Functionalities  & \# Instances & \# Traces & \# Metrics \\ \midrule
Dataset $\mathcal{A}$ & 292  & 2,035  &  100.2 M &  7.25 M \\ \midrule
Dataset $\mathcal{B}$ & 105 & 1,027 & 121.6 M & 3.71 M\\ \midrule
Total &397 & 3,062 & 212.6 M & 10.96 M \\ \bottomrule
\vspace{-5pt}
\end{tabular}
\end{table}

\subsection{Experimental Setup}\label{sec: exp_setup}

\noindent\textbf{Dataset.}
We evaluate \nm using two datasets collected from the production environment of \cloud.
To evaluate the generalizability of \nm, the two datasets ($\mathcal{A}$ and $\mathcal{B}$) are collected from two different geographically isolated regions with different numbers of users.
The detailed statistics of the two datasets are listed in Table~\ref{tab: dataset_statistics}.
These datasets only include instances that are subscribed by internal customers, where we are able to manually inspect their functionalities by collaborating with corresponding teams.
We select the instances running on our production environment that are most frequently invoked according to their communication traces.
Then, we reach the owners of these instances to figure out the concrete functionalities these instances support, and we finally obtain 3,062 labeled instances.
Although we are unable to fully cover all instances within the \cloud due to the manual effort required, our datasets encompass a diverse range of functionalities (397 types in total), such as databases, disaggregated memory, authentication servers, search engines, and machine learning algorithms. 
% \zb{need to mention that these functionalities could belong to different applications. Our goal is to figure out how each application deploys their instances with these functionalities.}\jy{changed the following:}
Such diversity would help evaluate whether a clustering algorithm can generalize to different functionalities. 
Additionally, these functionalities can belong to different applications. For example, while various applications may each have their own databases, these database functionalities are distinguished from one another in our datasets since they are utilized by distinct applications that serve diverse workloads (e.g., databases of an online shopping application and a face recognition application).
For the monitoring data, traces are extracted from the network packet transmission records, while metrics are collected at five-minute intervals. Given the extensive usage and frequent communication of instances, we ultimately collect hundreds of millions of traces. In terms of metrics, the total number of points is 10.96 million for all instances.
We have made our datasets publicly available in our GitHub repository. However, due to confidentiality concerns, the actual functionality names have been anonymized and are represented as ``cluster\_ID''.

% \noindent\textbf{Implementation Details.}
% We conducted the following experiments on a Linux server running Ubuntu 16.04.7 LTS with 512G of RAM and 72 cores Intel(R) Xeon(R) Gold 6140 CPU @ 2.30 GHz.
% We perform a grid search to find the hyper-parameter setting that produces the best result for baseline methods. 
% We implement CloudCluster and ROCKA based on mature libraries (\eg scikit-learn~\cite{sklearn}). For OmniCluster, we directly use the source code released by the authors~\cite{omniclsuter_code}.

\noindent\textbf{Evaluation Metrics.}
We use the metrics \textit{homogeneity}, \textit{completeness} and \textit{V-measure} to evaluate the effectiveness of \nm in grouping the instances within the same \cluster.
These metrics have been widely adopted in evaluating the quality of clustering results in previous studies. 
Homogeneity measures the proportion of instances in the same cluster that share the same ground truth labels. 
Completeness, on the other hand, measures the proportion of instances with the same ground truth labels that are grouped into a single predicted cluster. 
V-measure is a harmonic mean of homogeneity and completeness, providing an overall indicator for clustering performance considering the trade-off between these two metrics.

\noindent\textbf{Competitors.}
We select the competitors from recent studies:

\begin{itemize}[leftmargin=*, topsep=0pt]
   
    \item \textit{OSImage} is a basic baseline that uses the name of the operating system (OS) image to differentiate between instances. Cloud providers offer various pre-installed OS images to cater to diverse customer needs. For example, an OS image named \textit{deeplearning-pytorch-2.0} implies that the instance is designed for executing deep learning applications.
    
    \item \textit{CloudCluster}~\cite{pang2022cloudcluster} clusters instances based on their pairwise traffic matrix in cloud projects to determine the functional structure of the cloud service. It normalizes each row of the traffic matrix by feature scaling, then reduces its dimensionality through low-rank approximation. Finally, HCA is employed to group all instances.

    \item \textit{ROCKA}~\cite{li2018robust} aims to cluster instances by using their monitoring metrics. ROCKA first normalizes the metrics to eliminate amplitude differences. It then uses shape-based distance (SBD) as a distance measure, which is robust to phase shift and efficient for high-dimensional time series data. Then, clusters are created based on DBSCAN algorithm.
    % based on the shape similarities of baselines using 

    \item \textit{OmniCluster}~\cite{zhang2022robust} clusters instances based on multivariate metrics of each instance. It employs a one-dimensional convolutional autoencoder (1D-CAE) to extract the low-dimensional features of all metrics. These features are selected based on their periodicity and redundancy. Finally, it uses HAC to divide all instances into different clusters.
\end{itemize}

%%%%%%%%%%%%%%%%%%%%%%%%%%%%%%%%%%%%%%%%%%%%%%%%%%%%
\begin{table}[]
\footnotesize
\centering
\label{tab: rq1_effectiveness}
\caption{Effectiveness of \cluster Discovery}
\vspace{-5pt}
  \resizebox{0.5\textwidth}{!}{%
\begin{tabular}{ccccccc}
\toprule
\multirow{2}{*}{Methods} & \multicolumn{3}{c}{Dataset $\mathcal{A}$} & \multicolumn{3}{c}{Dataset $\mathcal{B}$} \\ %\cline{2-7} 
 & Homo. & Comp. & V Meas. & Homo. & Comp. & V Meas. \\ \midrule
OSImage &  0.238 & 0.894 & 0.376 & 0.258 & 0.889 & 0.400 \\
CloudCluster & 0.346 & 0.748 & 0.473 & 0.369 & 0.753 & 0.495 \\ 
ROCKA & 0.831 & 0.882 & 0.856 & 0.875 & 0.900 & 0.887  \\ 
OmniCluster & 0.932 & 0.862 & \underline{0.896} & 0.944 & 0.877 & \underline{0.909}  \\ \midrule
\nm & 0.976 & 0.916 & \textbf{0.945} & 0.979 & 0.922 & \textbf{0.950}  \\
 % $\Delta$(\%)  & +1.88\% & +0.99\% & +10.40\% & +3.38\% & +2.39\% & +7.10\% \\
 \bottomrule
\end{tabular}%
}
\end{table}
%%%%%%%%%%%%%%%%%%%%%%%%%%%%%%%%%%%%%%%%%%%%%%%%%%%%

\subsection{\textbf{RQ1:} Effectiveness in \cluster Discovery}

In this RQ, we evaluate the accuracy of the \clusters discovered by \nm in comparison with state-of-the-art baseline methods.
To achieve this, we apply \nm and baseline methods to cluster instances in the dataset of \A and \B.
We present the results of our experiments in terms of homogeneity (Homo.), completeness (Comp.), and v-measure (V Meas.) in Table~\ref{tab: rq1_effectiveness}, where we highlight the best V Meas. with boldface and the second-best ones with underline. 

% \nm outperforms others
It can be observed that \nm outperforms three state-of-the-art baseline methods, namely CloudCluster, ROCKA, and OmniCluster, by a significant margin, achieving V-measures of 0.945 and 0.950 on datasets \A and \B, respectively. These results indicate that \nm can achieve the best balance between homogeneity and completeness. 
This can be attributed to the fact that \nm effectively integrates communication and resource usage patterns to discover \clusters. 
Unlike \nm, baseline methods typically focus on either trace or metric data, leading to worser performance. Specifically, OSImage exhibits low homogeneity but high completeness, as using only image names to separate instances can overly group instances with different functionalities that share the same images. While CloudCluster outperforms OSImage in v-measure, it falls short of other metric-using baseline methods, suggesting that metric similarities are more effective in distinguishing functionalities than communication trace similarities.

{
\vspace{-5pt}
\begin{tcolorbox}[breakable,width=\linewidth-2pt,boxrule=0pt,top=1pt, bottom=0pt, left=1pt,right=1pt, colback=gray!20,colframe=gray!20]
\textbf{Answer to RQ1:} 
\nm outperforms all state-of-the-art comparative methods in revealing the \clusters across two different datasets, achieving a v-measure of 0.945 and 0.950 in dataset \A and \B. 

% This indicates \nm effectively integrate resource usage patterns and communication patterns to identify the underlining \clusters.

% the most superior solution that effectively clusters instances performing the same functionalities, with 
\end{tcolorbox}
\vspace{-10pt}
}

%%%%%%%%%%%%%%%%%%%%%%%%%%%%%%%%%%%%%%%%%%%%%%%%%%%%
\begin{table}[]
\footnotesize
% \small
\centering
\label{tab: rq2_ablation}
\caption{Contribution of Different Components in \nm}
\vspace{-5pt}
  \resizebox{0.5\textwidth}{!}{%
\begin{tabular}{ccccccc}
\toprule
\multirow{2}{*}{Methods} & \multicolumn{3}{c}{Dataset $\mathcal{A}$} & \multicolumn{3}{c}{Dataset $\mathcal{B}$} \\ %\cline{2-7} 
 & Homo. & Comp. & V Meas. & Homo. & Comp. & V Meas. \\ \midrule
 \nm & 0.976 & 0.916 & \textbf{0.945} & 0.979 & 0.922 & \textbf{0.950}  \\
\nm w/o Metrics & 0.462 & \textbf{0.920} & 0.615 & 0.463 & \textbf{0.949} & 0.622   \\ 
\nm w/o Traces & 0.949 & 0.869 & \underline{0.907} & 0.915 & 0.893 & \underline{0.904}  \\ \bottomrule
\end{tabular}%
}
\end{table}
%%%%%%%%%%%%%%%%%%%%%%%%%%%%%%%%%%%%%%%%%%%%%%%%%%%%

\subsection{\textbf{RQ2:} Contribution of Each Component~\label{sec: ablation_study}}
% \nm consists of two core components \ie, trace-based partitioning and metric-based clustering. 
In this RQ, we evaluate each component's contribution to \nm's overall performance. We created two \nm variants and compared them with the original approach across datasets \A and \B. The first, \textit{\nm w/o metrics}, eliminates metric-based clustering, relying solely on communication destination similarity. The second, \textit{\nm w/o traces}, omits trace-based partitioning, directly applying the HAC algorithm to cluster instances based on resource usage patterns.
% \begin{itemize}[leftmargin=*, topsep=0pt]
%     \item . 
%     \item 
% \end{itemize}

We present the comparison results in Table~\ref{tab: rq2_ablation}, from which we make the following observations. (1) Removing either of the two components can adversely affect the performance of \nm, underscoring the necessity of integrating both communication and resource usage patterns. 
(2) The V-measure of \textit{\nm w/o metrics} is significantly lower than that of \textit{\nm} and \textit{\nm w/o traces}, primarily due to its low homogeneity. This suggests that the trace-based partitioning step over-clusters many instances that should be separated.
The communication pattern alone is not distinctive enough because instances having different functionalities should still communicate with some common instances, such as network gateway and proxy services (as illustrated in Fig.~\ref{fig: pattern_study}-(a)).
Nonetheless, the use of solely communication patterns achieves the best completeness score, implying that it barely separates clusters that should be grouped.
(3) \textit{\nm w/o traces} has the lowest completeness score, indicating that it can overly split clusters apart, but it has a considerably high homogeneity.
This observation implies that \nm harnesses the benefits of both performance metrics and communication traces, achieving the optimal balance between homogeneity and completeness.

{
\vspace{-5pt}
\begin{tcolorbox}[breakable,width=\linewidth-2pt,boxrule=0pt,top=1pt, bottom=0pt, left=1pt,right=1pt, colback=gray!20,colframe=gray!20]
\textbf{Answer to RQ2:} 
The variants, \textit{\nm w/o metrics} and \textit{\nm w/o traces}, each sacrifice either homogeneity or completeness. Yet, \nm effectively combines communication traces and metric data, yielding the highest v-measure, \ie a balanced performance in completeness and homogeneity.
\end{tcolorbox}
\vspace{-8pt}
}

\subsection{\textbf{RQ3:} Parameter Sensitivity~\label{sec: hyperparameter_study}}
In the design of \nm, we identify the following two parameters that are manually selected and potentially affect the performance of \nm. For clarity, we present the evaluation results in Dataset \B; similar results are obtained in dataset \A.

% \zb{recall that the datasets contain different applications, so \nm is generalizable}
% \zb{talk about different applications}

\subsubsection{LSH threshold ($\theta_{LSH}$)} 
In \S\ref{sec: trace_based_partition}, we utilize LSH algorithm to perform a search for similar neighbors during the trace-based partitioning step. 
The LSH algorithm groups similar items together into the same bucket with high probability, but it cannot guarantee that all items in the same bucket are actually similar; therefore, $\theta_{LSH}$ is utilized to filter dissimilar items within each bucket.

\begin{figure}[t]
%   Impact of alerts with different levels of severity
    \subfigure[Impact of $\theta_{LSH}$ \label{fig: lsh_threshold}]
    {
        \includegraphics[width=0.465\linewidth]{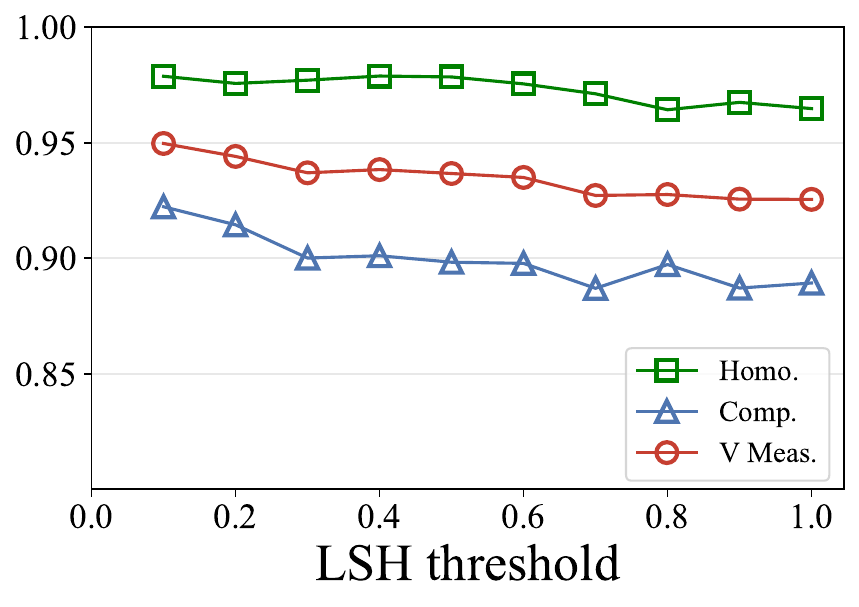}
    }
    \hspace{-16pt}
    \quad
    \subfigure[Impact of $\theta_{HAC}$ \label{fig: hac_threshold}]
    {
        \includegraphics[width=0.465\linewidth]{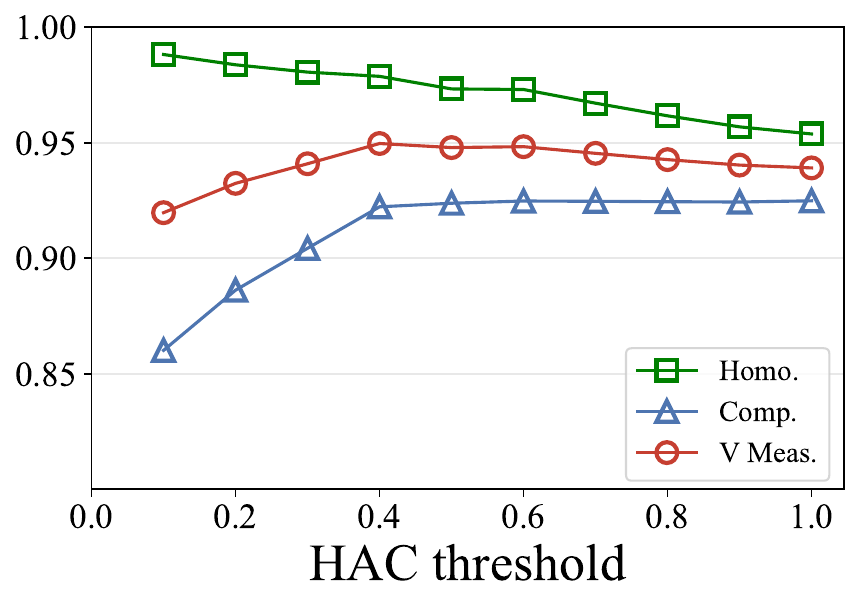}
    }
    \vspace{-4pt}
  \caption{Parameter Sensitivity of \nm}
\end{figure}

We varied the value of $\theta_{LSH}$ from 0 to 1 with a step size of 0.1 and evaluated the performance of \nm. The results, shown in Fig.~\ref{fig: lsh_threshold}, indicate that the V-measure remains stable with only a slight decrease as $\theta_{LSH}$ increases, which is primarily due to the decrease in completeness. This is because the LSH algorithm has already grouped similar items together into different buckets. Furthermore, since the communication patterns of most instances are distinct from one another, as depicted in Fig.~\ref{fig: pattern_study}-(a), there are only a small number of dissimilar items in the same bucket. As a result, adjusting $\theta_{LSH}$ does not significantly affect the clustering results.

\subsubsection{HAC threshold ($\theta_{HAC}$)}
In \S\ref{sec: metric_based_clustering}, HAC is used for clustering instances within each chunk, where the parameter $\theta_{HAC}$ controls the granularity of clustering: a smaller value of results in more fine-grained clusters, while a larger value results in fewer, coarser clusters.

We enumerated the value of $\theta_{HAC}$ from 0 to 1 with a step size of 0.1 and evaluated the performance of \nm. The results are shown in Fig.~\ref{fig: hac_threshold}. We observed that increasing $\theta_{HAC}$ can increase completeness and decrease the homogeneity. This is because larger clusters are generated when $\theta_{HAC}$ is larger.
The best v-measure is achieved when $\theta_{HAC}$ is around 0.4.
Subsequently, there is a slight decrease in homogeneity, while the v-measure remained stable. This decline in homogeneity is due to the inclusion of more dissimilar instances in a cluster, thereby reducing its homogeneity. Nevertheless, the preceding trace-based partitioning step groups similar instances together, resulting in a limited number of dissimilar instances. Hence, the overall performance is not significantly affected.

{
\vspace{-5pt}
\begin{tcolorbox}[breakable,width=\linewidth-2pt,boxrule=0pt,top=1pt, bottom=0pt, left=1pt,right=1pt, colback=gray!20,colframe=gray!20]
\textbf{Answer to RQ3:} 
\nm is not significantly sensitive to the parameters $\theta_{LSH}$ and $\theta_{HAC}$. This is because the trace-based partitioning step already groups similar instances together and separates dissimilar instances based on their communication patterns. Thus, adjusting these two parameters only has a minor effect on the clustering results.
% Considering that the dataset used contains different applications with various functionalities, it also indicates that \nm is robust to threshold selections and generalizable to identify functional clusters with a diverse range of patterns.
\end{tcolorbox}
\vspace{-8pt}
}

\subsection{\textbf{RQ4:} Efficiency of \nm}\label{sec: rq4_efficiency}
In this section, we assess the efficiency of \nm in the context of large-scale cloud systems with millions of instances that are frequently created, deleted or updated. 
% These instances may be frequently created, deleted or updated, so it is crucial to have an efficient solution that can handle such massive data and execute frequently within a reasonable timeframe, such as 24 hours.
To this end, we apply them to 1,000 / 5,000 / 10,000 / 50,000 / 100,000 instances and record the time needed (in seconds) to complete the clustering process. 
% Note that these instances are unlabeled since it is labor-sensitive to obtain their ground-truth functionalities.

Table~\ref{tab: efficiency} presents the results, from which we can make the following observations:
(1) ROCKA, OmniCluster, and \nm w/o Traces require increasingly more time as the number of instances increases, and they cannot complete the clustering process within a reasonable time when clustering 100,000 instances. This is mainly because these methods require pair-wise similarity computation based on instance metrics, resulting in a quadratic growth in time complexity as the number of instances increases. OmniCluster mitigates this issue by reducing the dimensionality of metrics, requiring less time than the other two methods.
(2) CloudCluster and \nm w/o Metrics are more efficient than other baseline methods. \nm w/o Metrics is more efficient because we optimize efficiency using pair-wise comparison with LSH and MinHash, as described in \S\ref{sec: trace_based_partition}.
(3) \nm is less efficient than \nm w/o Metrics since it requires an additional metric-based clustering step.
In addition, when the number of instances is fewer than 10,000, CloudCluster outperforms \nm because the time required by \nm to build the LSH index is dominant. However, as the number of instances increases to 100,000, \nm's efficiency becomes superior to other baselines, being four times faster than CloudCluster. 
This is attributed to the coarse-to-fine clustering process, which limits pairwise distance computation within small chunks.
Therefore, the time cost of \nm only increases linearly with the instance numbers.

%%%%%%%%%%%%%%%%%%%%%%%%%%%%%%%%%%%%%%%%%%%%%%%%%%%%

%%%
% May be 1000, 5000, 10000, 50000, 100000?

\begin{table}[]
\footnotesize
\centering
\label{tab: efficiency}
\caption{Efficiency Comparison with Increasing Scales of Instances}
% \vspace{-4pt}
  \resizebox{0.47\textwidth}{!}{%
\begin{tabular}{cccccc}
\toprule
\multirow{2}{*}{Methods} & \multicolumn{5}{c}{\# Instances} \\ %\cline{2-7} 
 & 1,000  & 5,000 & 10,000 & 50,000 & 100,000 \\ \midrule
CloudCluster & 0.9 & 23.87 & 78.65 & 1768.7 & 5585.7 \\ 
ROCKA & 80.7 & 1981.8 &  7850.3 & - & - \\ 
OmniCluster & 31.7 & 264.6 & 1048.6 & 26531.8 &  - \\ \midrule
\nm w/o Metrics & 3.9 & 19.1 & 40.2 & 195.1  &  392.4  \\ 
\nm w/o Traces  & 80.3 & 2066.1  & 8232.3 & -  & - \\  \midrule
 \nm & 18.2 & 89.4 & 183.9 & 929.2 & 1912.7 \\ \bottomrule

\end{tabular}%
}
\end{table}
%%%%%%%%%%%%%%%%%%%%%%%%%%%%%%%%%%%%%%%%%%%%%%%%%%%%

{
% \vspace{-5pt}
\begin{tcolorbox}[breakable,width=\linewidth-2pt,boxrule=0pt,top=1pt, bottom=0pt, left=1pt,right=1pt, colback=gray!20,colframe=gray!20]
\textbf{Answer to RQ4:} 
Compared with state-of-the-art solutions, \nm is the most efficient solution when processing a large number of instances (\eg 100,000).
Moreover, thanks to the coarse-to-fine strategy of \nm, its time cost increases linearly with an increasing number of instances, making it scalable for handling massive instances in cloud systems.
\end{tcolorbox}
% \vspace{-8pt}
}

%% file: content/05_casestudy.tex
\section{Industrial Experience}\label{sec: case_study}
In this section, we share our experience in applying \nm to a real-world cloud system (\ie \cloud), which aims to demonstrate the practical usefulness of \nm. 
Generally, customers usually subscribe instances from \cloud in a batch manner, \eg thousands of instances.
These customers can then concentrate on the development and deployment of a variety of services across these instances, while the cloud providers handle the often tedious tasks of maintenance and operation to ensure system reliability.
Due to privacy concerns, on-site engineers from \cloud can only rely on limited runtime information of these instances (\eg network packet drop rate) to monitor their health states~\cite{chen2020incidental,liu2023incident,li2021fighting}.
However, without knowing how customers' applications are organized in these instances, we observe that some potential threats in the deployment or underlining errors may be missed, which may later cause service interruptions, consequently impacting the overall availability of the deployed applications~\cite{pang2022cloudcluster}.
To address this problem, in \cloud, we adopt \nm to reveal \clusters in the massive instances hosted by \cloud. 
These functional clusters provide additional information regarding the structure of service deployment across the instances, thus enabling us to conduct more comprehensive and fine-grained monitoring of the cloud system.
We present two primary usage scenarios of functional clusters within \cloud: \textit{vulnerable deployment identification} and \textit{latent issue discovery}.
% \begin{figure}[t]
%     \centering
%     \includegraphics[width=1\linewidth]{figures/vulnerable_deployment.pdf}
    
%     \caption{Case I: latent issue discovery}
%     \label{fig: latent_issue_case}
% \end{figure}

% \begin{figure}[t]
% %   Impact of alerts with different levels of severity
%     \subfigure[ \label{fig: latent_issue_case_all}]
%     {
%         \includegraphics[width=0.4\linewidth]{figures/all_broken.pdf}
%     }
%     \hspace{-16pt}
%     \quad
%     \subfigure[ \label{fig: latent_issue_case_part}]
%     {
%         \includegraphics[width=0.55\linewidth]{figures/app_broken.pdf}
%     }
    
%   \caption{Case I: latent issue discovery}
% \end{figure}

\begin{figure}[t]
    \centering
    \includegraphics[width=1\linewidth]{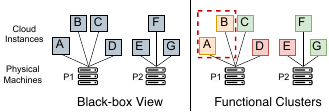}
    % \vspace{-8pt}
    \caption{Case I: vulnerable deployment identification}
    \label{fig: vulnerable_deployment_case}
\end{figure}

\subsection{Vulnerable Deployment Identification}
Functional clusters can help cloud providers identify instances with vulnerable deployments. 
Specifically, a vulnerable deployment refers to a scenario where all instances, having the same functionalities, are deployed on the same physical machines.
In such case, once a failure happens on this physical machine (\eg disk failure~\cite{liu2019bugs}), the entire functionality can be interrupted. 
In contrast, if these instances were distributed across different physical machines, only a subset of the instances would be affected in the event of a failure, thereby preventing a complete shutdown of the functionality.
However, due to the abstraction of physical resources into instances, customers often deploy their applications within these instances without understanding how these instances are distributed across actual physical machines. 
On the other hand, cloud vendors possess knowledge of the mapping between instances and physical machines; yet, they are often unaware of the organization of functionalities across these instances due to privacy concerns.
% \jz{cloud vendors are aware of the instance deployment but without information about the functionalities due to...}.
Given the vast number of instances in a cloud system, manually identifying vulnerable deployments poses a significant challenge for on-site engineers.

To fill in this gap, we apply \nm to identify functional clusters to help detect potentially vulnerable deployments.
Fig.~\ref{fig: vulnerable_deployment_case} provides a concrete example. The left-hand side presents a black-box view of instance deployment from a cloud vendor's perspective, where only the information about which instances are deployed on which physical machines is known.
In contrast, the right-hand side displays instances with \clusters. With this knowledge, we can identify three functionalities: a functionality including A and B (marked as yellow), a functionality including D and E (marked as red), and a functionality including C, F, and G (marked as green). 
The deployment of the yellow functionality is potentially vulnerable because both A and B are deployed on physical machine P1. 
In contrast, the other two functionalities are more reliable since their instances are deployed across two different physical machines, making them resilient to the failure of either machine. 
% 看一下加这一段是否合适@JY @ZB
It is worth noting that although \nm can hardly pinpoint what specific functionality of an instance serves, it can identify the instance group having the same functionalities, which facilitates automatic vulnerable deployment identification without violating privacy policies.

We have applied \nm in \cloud to discover functional clusters for around 3,000 internal instances and identified \textit{eight} cloud services with vulnerable deployments. 
We then contacted the corresponding teams, confirmed the existence of the vulnerable deployments, and assisted in migrating the instances across different physical machines for improved resilience.
In the future, our goal is to broaden the adoption of \nm to benefit a wider group of users and help enhance the reliability of their application deployment.

% \jy{triggered by operators to identify possible vulnerable deployment}

\begin{figure}[t]
    \centering
    \includegraphics[width=1\linewidth]{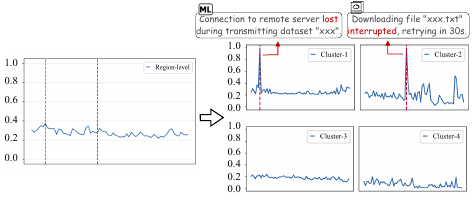}
    \vspace{-20pt}
    \caption{Case II: latent issue discovery}
    \label{fig: latent_issues_case}
\end{figure}

\subsection{Latent Issue Discovery}
% \clusters provide us with an application-grain view to discover more significant data characteristics from a vast amount of data generated by these instances. 
The second typical use case of \nm in \cloud is to identify latent network issues that may not be discovered by traditional monitoring methods.
Modern cloud providers have been equipped with various monitoring tools to ensure the quality of their network services (\eg flow logging of AWS~\cite{aws_flowlog}). 
% Monitoring network quality is one crucial part of the monitoring system.
It is essential for such monitoring tools to comprehensively discover underlining problems in the cloud systems that can affect user experience, but without firing too many false alarms to distract the on-site engineers.

One crucial type of network monitoring is to monitor the packet loss of each instance.
Packet loss, which denotes network packets that are accidentally dropped, can usually occur in any instance of a cloud system.
However, they may not necessarily indicate a problem, as these errors could be caused by transient network congestion and may not affect users' experience.
Considering the vast number of instances in a large-scale cloud system, a significant number of packets could be lost every minute. 
This presents a challenge for cloud vendors in converting this fragmented packet loss data into actionable alarms for on-site engineers.

To address this problem, we resort to the aggregation of packet loss data from a selected group of instances, using an appropriate granular approach to identify potential problems.
The underlying assumption here is that \textit{simultaneous} packet losses occurring within a group of instances are more likely to impact user applications.
For instance, if all instances within a region experience packet loss within a short time frame, it strongly suggests a regional network issue.
However, one large region can contain millions of instances, and consequently, grouping by a region might fail to reveal local issues for a particular application.
Another possible solution is to utilize the metadata (\eg the TenantID of the customer) to group instances.
Nonetheless, there could still be tens of thousands of instances associated with the same identifiers~\cite{pang2022cloudcluster}. For example, all instances subscribed to by the same enterprise customer would share the same identifier.

\nm enables a more effective approach, which is to aggregate lost packets in the granularity of the (approximated) \clusters, which can reveal latent issues that may not be visible at neither a coarser level (\eg regional level) nor a finer level (\eg instance level).
Fig.~\ref{fig: latent_issues_case} shows the changes in the number of lost packets (normalized) calculated at either the region grain (left-hand side) or functionality grain (right-hand side). 
We can observe that while the numbers of packet loss barely change for the whole region, some functionalities (\ie Cluster-1 and Cluster-2) experience sudden increases in packet loss.
This indicates that there may be latent issues affecting the performance of those specific functionalities, which are unnoticed if monitored at the region level. 
We then contact the corresponding teams and confirm that Cluster-1 and Cluster-2 correspond to machine learning and storage functionalities, respectively.
We then validate these latent issues, and both functionalities experience interruption due to unstable network states, as evidenced in their log messages shown in Fig.~\ref{fig: latent_issues_case}.
This highlights the potential of \nm in facilitating identifying issues that customers are experiencing without accessing their private data, which allows cloud vendors to provide more comprehensive monitoring to enhance the reliability of the cloud systems.
% \vspace{-6pt}

\textbf{Enhanced Cloud Monitoring Based on \nm.}
To summarize, these two use cases demonstrate that \clusters can be utilized with existing monitoring tools and enable identifying vulnerable deployment and discovering latent issues automatically.
\nm plays a crucial role to provide comprehensive and precise \clusters for large-scale instances. 
With the significant growth of modern cloud systems, instances experience frequent dynamic changes, including creation, deletion, and migration. In this context, \nm can be utilized to efficiently capture relations between instances. 
Unlike using pre-defined and rule-based monitoring~\cite{li2021fighting,gu2020efficient}, \nm is adaptive to the frequent evolution of cloud applications.
By continuously monitoring metrics like packet loss and the distribution of instance deployments, the monitoring system can effectively detect anomalies, such as sudden spikes in packet loss or scenarios indicating vulnerable deployments.
This enables prompt alerts to the on-site engineers of relevant teams, resulting in shorter response time and more efficient issue resolution. Overall, the effectiveness and efficiency of \nm significantly contribute to improving the overall monitoring and management of instances in modern cloud systems.

 % \jz{In addition, unlike pre-defined, static monitoring, \nm is adaptive to the frequent evolution of cloud applications, \ie the \clusters can change ... while \nm can ...}
 
% As shown in  (left-hand side), which displays in a region of the cloud over a day, there may not be any significant network problems occurring.
% In contrast,  as shown in Fig.~\ref{fig: latent_issue_case} (right-hand side). 
% This provides a more proper grain of the data, .
% As a result, it is hard to determine when 

%% file: content/06_discussion.tex
% \section{Discussion}
\section{Threats to Validity}

\textbf{External Validity.} The primary external threat of this study is the investigated object.
The datasets are collected from \cloud, as there are no publicly available datasets that include both instance data and corresponding functionality labels.
However, \cloud is a world-leading cloud provider with a vast scale. The data collected from the production environment records real behaviors of instances and covers a broad range of functionalities from two large regions as detailed in \S\ref{sec: exp_setup}. 
Therefore, the \cloud evaluation is representative and convincing. The data used by \nm, which includes traces and metrics, is typically collected by modern cloud vendors like AWS~\cite{aws_cloudwatch} and GCP~\cite{gcp_monitoring}. This suggests that our solution could be applied to similar cloud systems, potentially benefiting cloud customers globally.
% We also open-source the datasets after removing sensitive information to further benefit similar studies in the community.

\textbf{Internal Validity.} The primary internal factors that could potentially compromise validity are implementation and parameter setting.  To address the implementation threat, we closely followed the original papers for baseline approaches that lacked open-sourced code and re-implemented them accordingly. To minimize this threat further, we utilized several mature libraries (\eg scikit-learn) for implementing the core algorithms.  Moreover, both our proposed methods and the baseline methods were subject to rigorous peer code review.
To mitigate the parameter setting threat, we fine-tuned the baseline methods utilizing a grid-search approach, subsequently selecting the most optimal results.

% For our proposed method \nm, we conducted a comprehensive evaluation to assess the impact of parameter selection, as elaborated in \S\ref{sec: hyperparameter_study}.

% \vspace{-6pt}

%% file: content/07_relatedwork.tex
\section{Related Work}

% Our study is inspired by studies concerning instance clustering and reliability in cloud systems summarized as follows.

\subsection{Instance Clustering}

% Cloud providers struggle to manage millions of instances in cloud systems. 
% Cloud vendors mainly utilize two data sources (\ie communication traces and performance metrics) to gain more insights about these instances for improved management.
Communication traces are usually modeled as a communication graph for clustering instances. 
Xu et al~\cite{xu2011network} perform network-aware clustering for end hosts with the same network prefixes by using bipartite communication graphs. \cite{iliofotou2007network}\cite{jin2009unveiling}\cite{nagaraja2010botgrep} attempt to mine the pattern of instance-to-instance communication, then detect abnormal traces to safeguard the instances.
Pang et al.~\cite{pang2022cloudcluster} propose CloudCluster, which uses a novel combination of feature scaling, dimensionality reduction, and hierarchical clustering to cluster a large scale of instances.
Another line of work utilizes various technologies to model multivariate metric data of instances.
For example, Kane et al.~\cite{kane2017multivariate} employ Principal Component Analysis (PCA) to transform multivarite metric data to univariate time series before clustering. 
The most recent work, ROCKA~\cite{li2018robust} adopts shape-based distance (SBD)~\cite{paparrizos2015k} as a robust distance measurement for clustering.
OmniCluster~\cite{zhang2022robust} adopts hierarchical agglomerative clustering to cluster instances represented by low-dimension representations. 
Contrary to previous studies, \nm effectively combines communication traces and multivariate metric data, surpassing state-of-the-art solutions by utilizing both data types, as shown in Section~\ref{sec: evaluation}.

% However, using only MTS of KPI data is not ideal for large-scale cloud systems, since there can be thousands of VMs and each VM generates a large amount of KPI data. Most of the methods need to compute distance metrics between VMs, and the time and space complexity are unacceptable when the number of VMs is large. In addition, the deep learning based clustering methods lack of interpretability and are sensitive to noise and parameter settings, and they usually require long training time. These drawbacks make them not practical in real scenarios. In contrast, by leveraging traffic flow data, \cluster can initially divide VMs into coarser-grained chunks, thus reducing the overhead of clustering.

\subsection{Reliability of Cloud Systems}
% Modern cloud systems are becoming increasingly large-scale and complex to understand, which poses challenges to ensuring their reliability.
Extensive efforts have been made to examine and understand the important factors contributing to cloud system reliability.
For example, Chen et al.\cite{chen2020towards} extensively studied large-scale public cloud incidents, analyzing disruptions and failures to pinpoint reliability issues. Similarly, Huang et al.\cite{huang2017gray} explored the effects of gray failures in cloud systems.
In addition, other studies~\cite{gunawi2016does,liu2019bugs,cotroneo2019bad,ghosh2022fight} reviewed public or internal postmortem reports and summarized the causes of outages in cloud systems.
These studies underscore the need for enhanced understanding and observability of interdependent cloud system components.
Furthermore, researchers have explored automated solutions for timely incident detection~\cite{gu2020efficient,zhao2020understanding,li2021fighting,chen2020identifying} and failure mitigation~\cite{chen2019outage, wang2021fast, liu2023incident, chen2022online, chen2021graph} in cloud systems.
Despite the promising results of these solutions in improving cloud system reliability, most~\cite{wang2021fast, chen2020identifying, chen2021graph} rely on well-abstracted incident descriptions and clear system topologies, often unavailable or incomplete to cloud providers~\cite{liu2023incident}. In contrast, \nm's \clusters can supplement these methods and provide insights to enhance these tasks, as shown in \S\ref{sec: case_study}.

%% file: content/08_conclusion.tex
\section{Conclusion}

This paper presents an approach to enhance the observability of cloud systems by inferring functional clusters of instances. 
To achieve this, we conduct a pilot study based on the real-world datasets collected in \cloud, indicating that communication patterns and resource usage patterns are two essential indicators for revealing functional clusters.
Motivated by our findings, we propose a non-intrusive, coarse-to-fine clustering method, \nm, which effectively integrates both communication and resource usage patterns. 
Experiments on two industrial datasets are conducted to evaluate \nm.
Our results show that \nm outperforms state-of-the-art solutions with a v-measure of 0.95; and \nm can efficiently process massive instances.
Furthermore, we share our experiences in applying \nm in \cloud.
Two cases, \ie vulnerable deployment identification and latent issue discovery, demonstrate the usefulness of \nm in improving the reliability of \cloud.

%% file: content/ack.tex
\section{Acknowledgement}
The work described in this paper was supported by the Key-Area Research and Development Program of Guangdong Province (No. 2020B010165002) and the Key Program of Fundamental Research from Shenzhen Science and Technology Innovation Commission (No. JCYJ20200109113403826). It was also supported by the Research Grants Council of the Hong Kong Special Administrative Region, China (No. CUHK 14206921 of the General Research Fund).
Thanks to Hongliang Xiang from the EI Public Service Domain Program at Huawei Cloud Computing Technologies Co., Ltd for his assistance with data collection and annotation.

% hongliang xiang should be mentioned.